# Privacy via the Johnson-Lindenstrauss Transform

Krishnaram Kenthapadi[†]  Aleksandra Korolova[‡]
Microsoft Research  Stanford University

Ilya Mironov[†]  Nina Mishra[†]
Microsoft Research  Microsoft Research

[†]{krisken,mironov,ninam}@microsoft.com  [‡]{korolova@cs.stanford.edu}

April 13, 2012

**Abstract**

Suppose that party A collects private information about its users, where each user's data is represented as a bit vector. Suppose that party B has a proprietary data mining algorithm that requires estimating the distance between users, such as clustering or nearest neighbors. We ask if it is possible for party A to publish some information about each user so that B can estimate the distance between users without being able to infer any private bit of a user. Our method involves projecting each user's representation into a random, lower-dimensional space via a sparse Johnson-Lindenstrauss transform and then adding Gaussian noise to each entry of the lower-dimensional representation. We show that the method preserves differential privacy—where the more privacy is desired, the larger the variance of the Gaussian noise. Further, we show how to approximate the true distances between users via only the lower-dimensional, perturbed data. Finally, we consider other perturbation methods such as randomized response and draw comparisons to sketch-based methods. While the goal of releasing user-specific data to third parties is more broad than preserving distances, this work shows that distance computations with privacy is an achievable goal.

## 1 Introduction

In recent years, there has been an abundance of rich and fine-grained data about individuals in domains such as healthcare, finance, retail, web search and social networks. It is desirable for data collectors to enable third parties to perform complex data mining applications over such data. However, privacy is a natural obstacle that arises when sharing data about individuals with third parties, since the data about each individual may contain private and sensitive information.

We ask the following question: Is it possible to empower third parties with knowledge about users without compromising privacy of the users? Suppose that party A collects private information about its users, where each user's data is represented as a bit vector. We focus on the setting where a third party B has a proprietary data mining algorithm that requires estimating the distance between users, such as clustering or nearest neighbors. We ask if it is possible for party A to publish some information about each user so that B can estimate the distance between users without being able to infer any private bit of a user. Even in a scenario where the data mining algorithm is run by the data collector itself (that is, parties A and B are the same), privacy breaches are possible if the result of data mining is to be published [19]. The reason is that complex algorithms that access private data can be susceptible to either unintended or adversarial attacks [6]. While one way to address this problem is to design a sophisticated algorithm that respects privacy (e.g., [7]), our approach can ensure that the data given as input to the algorithm itself does not compromise user privacy.

Although estimating distances between users in a privacy-preserving manner is a fundamental primitive in many data mining applications, the approaches known to date have certain short-comings. Approaches resorting to user id anonymization while keeping the data unchanged have been shown to be badly insufficient to preserve privacy [2, 29]. A random projection based method to preserve distance between users was proposed in [22], but later work demonstrated concrete attacks to breach privacy on this method [14, 30]. These approaches suffer due to lack of a rigorous privacy definition. On the other hand, approaches to data sharing such as the recently proposed provably private methods for search query and click data release [16, 20] accomplish it at the price of giving up on all user-level information.

**Contributions:** We describe a simple, natural way to publish a sketch of a user that simultaneously preserves privacy and enables estimation of the distance between users. The main idea is to project a $d$-dimensional vector representation of a user's feature attributes into a lower $k$-dimensional space by first applying a random Johnson-Lindenstrauss transform and then adding Gaussian noise $N(0, \sigma^2)$ to each bit of the resulting vector. We prove that this perturbed lower-dimensional vector preserves differential privacy, i.e., an attacker who knows all but one attribute of a user cannot recover the value of that attribute from the published information with high confidence. In terms of utility, we show how to recover the distance between users from the perturbed sketches. We show that the squared Euclidean distance between pairs of users is preserved in expectation. Further, with high probability, the distance between users is preserved up to the usual Johnson-Lindenstrauss factors plus an additive factor that depends on $k$ and the variance of the noise $\sigma^2$.

We also compare our proposed solution to other candidate solutions. For instance, we compare to the more straightforward solution of directly adding noise to the user × user distance matrix. We show that in order to achieve the same privacy, the variance of the added noise is higher for the more direct method. Concretely, we show that the projection-based method is better if the maximal weight of a user vector is much smaller than the number of users. Also, we analyze a randomized response [32] method for data sharing. For a fixed value of the target dimension $k$, we show that for nearby points (those within squared distance $O(\sqrt{k})$), both algorithms are inaccurate. Projection-based methods are better when pairs are medium distance apart, i.e., between $\sqrt{k}$ and $\sqrt{dk}/\epsilon^2$. Randomized response methods excel for pairs that are far apart.

While the problem of sharing data with third parties is more complex than producing sketches of user data that preserve distances between users, this work offers a privacy-preserving method to enable third parties to execute one of the core data-mining primitives. Since the goal of our work is to enable distance computations, understandably it does not apply in situations and applications where proximity to a particular user is itself sensitive information.

## 2 Preliminaries

We first describe how the users are represented in our model and provide a formal problem statement. Then we discuss the measure of utility as well as the privacy definition. We also state a classic result on preserving distances during dimensionality reduction which is crucial for our techniques to work.

## 2.1 User representation

We represent each user belonging to a set $U$ of $n$ users as a binary vector in $d$ dimensions, where each dimension corresponds to the value of an attribute (e.g. gender, interest/disinterest in a particular topic, location information, etc.) We assume that the attribute meanings are not sensitive or, if they are, they can be published in a privacy-preserving manner (say, using the techniques in Göetz et al. or Korolova et al. [16, 20]).

Our goal can be formally stated as: *Given a set of user profiles represented as vectors in $d$ dimensions, publish sketches of the user profiles that simultaneously* **preserve user privacy** *and enable third parties to* **estimate pairwise distance between users**.

## 2.2 Utility measure

We consider Euclidean ($\ell_2$) distance between users as the distance measure we aim to preserve, as it is a natural choice for similarity search in high dimensions [15]. We discuss other distance measures in §5. Our measure of utility is whether pairwise Euclidean distance between users can be recovered by a third party, who has access only to the transformed privacy-preserving user profiles.

## 2.3 Privacy definition

Any system that employs heuristic notions of privacy suffers from the fundamental problem that an adversary can come up with sophisticated attacks to breach the protection in ways that the system designer had not anticipated. Hence we contend that it is crucial to design a data release method with provable privacy guarantees. We adopt a rigorous approach to privacy introduced by Dwork et al. [12], which has gained widespread recognition in recent years (see a survey [10]), and has been used to demonstrate the feasibility of privacy-preserving data releases [3, 20, 23, 24]. We adopt a slight variant of the definition introduced in Dwork et al. [11]:

**Definition 1** (($\epsilon, \delta$)-differential privacy). *A (randomized) algorithm $A$ satisfies ($\epsilon, \delta$)-differential privacy, if for all inputs $X$ and $X'$ differing in at most one user's one attribute value, and for all sets of possible outputs $\hat{D} \subseteq Range(A)$:*

$$\Pr[A(X) \in \hat{D}] \leq e^\epsilon \cdot \Pr[A(X') \in \hat{D}] + \delta,$$

*where the probability is computed over the random coin tosses of the algorithm.*

Intuitively, the differential privacy guarantee states that an attacker who knows all attributes of all users except one attribute of one user cannot infer with confidence the value of that attribute, from the information published by the algorithm. The $\delta$ parameter corresponds to the probability with which the preceding guarantee can fail, with $\delta$ typically thought of as $O(1/n)$. The privacy guarantees also extend to small collections of (not necessarily related) attributes.

The privacy guarantees may be achieved through introduction of noise to the output. In order to achieve the more stringent privacy guarantee of ($\epsilon, 0$)-differentially privacy, the noise added typically comes from the Laplace distribution [12]. If one is willing to tolerate a more lenient guarantee of ($\epsilon, \delta$), the noise can be added from the more tightly concentrated Normal distribution [11].

## 2.4 Johnson-Lindenstrauss transform

A celebrated result in geometry, the Johnson-Lindenstrauss Lemma [18], states that for any set $V$ of $n$ points in $\mathbb{R}^d$, given $\lambda_{\text{JL}} > 0$ and $k = \Omega\left(\frac{\log n}{\lambda_{\text{JL}}^2}\right)$, there exists a map that embeds the set into $\mathbb{R}^k$, distorting all pairwise distances within at most $1 \pm \lambda_{\text{JL}}$ factor. The proof proceeds by showing that for any $x, y \in \mathbb{R}^d$, a linear projection $P \in \mathbb{R}^{d \times k}$ sampled from a carefully defined class satisfies

$$(1 - \lambda_{\text{JL}})\|x - y\|_2^2 \leq \|xP - yP\|_2^2 \leq (1 + \lambda_{\text{JL}})\|x - y\|_2^2$$

with certain probability $1 - \delta_{\text{JL}}$ over the choice of the projection matrix, where $\log \frac{1}{\delta_{\text{JL}}} = O(k\lambda^2)$, and then applying the union bound.

This transform has become a fundamental tool in dimensionality reduction and similarity search in high dimensions, and computer science literature [1, 17] has proposed several constructions for $P$.

# 3 Construction and Usage of Privacy-Preserving Projections

We next describe the intuition as well as the technical components of our approach. Then we state our privacy and utility guarantees and provide their proofs.

## 3.1 Algorithms for transforming user profiles and recovering distances

Our mechanism for enabling data sharing with privacy consists of two components: 1) an algorithm that transforms the representation of each user into a privacy-preserving sketch and 2) an algorithm that recovers distances between users from the transformed user sketches. The intuition for the design of our mechanism is as follows: since we aim to preserve pairwise distances with the goal of performing user segmentation and nearest neighbor computations, an algorithm that performs a privacy-preserving transformation of user profiles while approximately maintaining pairwise distances would suffice. At the core of our method is a one-time privacy-preserving transformation of user profiles that can be published. All subsequent operations can be performed on this published data and therefore do not consume a "privacy budget" or pose additional privacy risk.

Our algorithms are easy to state and implement, and do not require understanding of privacy. However the proofs of privacy and utility guarantees are non-trivial and require deeper analysis.

### 3.1.1 Private projection algorithm

The goal of Algorithm 1 (PRIVATEPROJECTION) is to transform an $n \times d$ representation of user data into a representation that can be publicly shared without compromising the privacy of any individual involved and can simultaneously preserve distance characteristics of the original representation. First, the data is projected into a much lower dimension ($k \ll d$) to obtain a compact representation that preserves pairwise distances (steps 1–2), similar to many dimensionality reduction techniques. Then the resulting data is slightly perturbed (steps 3–4) to guarantee privacy of each user. The benefit of projecting onto a lower dimension and doing the perturbation is that we require less noise addition.

**Algorithm 1** PRIVATEPROJECTION

**Input:** Boolean $n \times d$ matrix $X$ whose rows correspond to people and columns correspond to attributes learned about the users by first parties; Privacy parameters $\epsilon, \delta$; Projected dimension $k$.
**Output:** $d \times k$ projection matrix $P$; privacy-preserving $n \times k$ matrix $Z$, both of which can be published.
1: Construct random $d \times k$ projection matrix $P$.
2: $Y := XP$
3: Construct random $n \times k$ noise matrix $\Delta$, based on privacy parameters $\epsilon, \delta$ and projection matrix $P$.
4: $Z := Y + \Delta$
5: Publish $(P, Z)$.

---

Intuitively, for a given level of desired privacy, there are two factors that affect utility and behave in opposite directions as we vary the projected dimension $k$. On the one hand, as $k$ gets smaller, dimensionality reduction plays a greater role in the distortion of distances. On the other hand, as $k$ gets larger, noise added plays a greater role in the distortion of distances. Finding the optimal value for the projected dimension $k$ is challenging theoretically, as it depends on the underlying data distributions and the specific distance values we are trying to preserve.

We next discuss the key components of Algorithm 1, namely, the choice of desired privacy guarantees $(\epsilon, \delta)$ which determines the distribution of the noise, the choice of projection matrix $P$ (and its *sensitivity*) and the corresponding choice of the parameters of the noise matrix's distribution. We remark that the projection matrix as well the noise matrix do not depend on $X$, but only require knowledge of the number of users $n$, the original dimension $d$ and the desired privacy parameters. Following Kerckhoffs's Principle in Cryptography, we assume that the algorithm as well as the parameters $n, d, k, \epsilon, \delta, P$ and the parameters used in the noise matrix are publicly known.

### 3.1.2 Choosing desired privacy guarantees

The first decision in utilizing Algorithm 1 (PRIVATEPROJECTION) is to determine the privacy guarantees desired by the algorithm's curator. The crucial observation is that one is able to guarantee $(\epsilon, \delta)$-differential privacy by adding noise $\Delta$ from the Normal distribution, with the variance of the noise depending on the $\ell_2$ sensitivity of the chosen projection matrix $P$, which we define next.

**Definition 2** ($\ell_\rho$-Sensitivity of $P$). *Define the $l_\rho$-sensitivity of a $d \times k$ projection matrix $P = \{P_{ij}\}_{d \times k}$, denoted by $w_\rho(P)$, as the maximum $\ell_\rho$-norm of any row in $P$, i.e., $w_\rho(P) = \max_{1 \leq i \leq d} \left(\sum_{j=1}^{k} |P_{ij}|^\rho\right)^{\frac{1}{\rho}}$.*

Equivalently, $w_\rho(P)$ can be defined as $\max_{e_i} \|e_i P\|_\rho$, where $\{e_i\}_{i=1}^{d}$ are standard basis unit vectors.

### 3.1.3 Choosing projection matrix $P$

There are many ways to choose a projection matrix for dimensionality reduction, depending on the properties of the data that need to be preserved. Our choice of $P$ is guided by two considerations: (1) we would like to preserve pairwise $\ell_2$ distances and thus user segmentation based on these distances (2) we would like to minimize the amount of noise to be added in order to maximize utility while guaranteeing privacy in the subsequent step.

The natural candidate projection matrices for (1), preserving $\ell_2$ distances between vectors, are the random projection matrices satisfying Johnson-Lindenstrauss guarantees (§2.4), such as the ones below:

1. Each entry of the matrix drawn independently from a Normal distribution with mean 0 and $\sigma^2 = 1/k$. [17].
2. Each entry of the matrix drawn independently and uniformly at random from $\{-\frac{1}{\sqrt{k}}, +\frac{1}{\sqrt{k}}\}$ [1].
3. Each entry of the matrix is chosen independently to be $+\sqrt{\frac{3}{k}}, 0, -\sqrt{\frac{3}{k}}$ with probability $\frac{1}{6}, \frac{2}{3}, \frac{1}{6}$, respectively [1].
4. The extremely sparse projection matrix of Dasgupta et al. [9].

As we will see in the proof of privacy in Theorem 1, when using noise $\Delta$ from Normal distribution, the amount of noise needed to preserve privacy depends on the choice of the projection matrix $P$; and more precisely, on the $\ell_2$-sensitivity of the chosen $P$. It is therefore desirable to use a projection matrix with low $\ell_2$ sensitivity, in order to ensure that we are adding the smallest possible amount of noise and therefore, maximizing utility while preserving privacy. The expected $\ell_2$ sensitivity of all of the random projection matrices described above is tightly concentrated around 1 (using the alternative definition of $w_2(P) = \max_{e_i} \|e_i P\|_2$, where $\{e_i\}_{i=1}^d$ are standard basis unit vectors, and by applying the proofs of low distortion for these matrices) and therefore, all of them are suitable for privacy preserving transformations that aim to preserve the maximum utility.

We emphasize that the specific measure of sensitivity of the matrix $P$, namely $\ell_2$-sensitivity, is driven by the type of noise added for privacy, which is Normal in our case, and not by the choice of norm one seeks to preserve under projection.

### 3.1.4 The random noise matrix $\Delta$

The choice of the desired privacy guarantees and projection matrix $P$ determines the noise matrix $\Delta$. Each entry in $\Delta$ is drawn randomly and independently from Normal distribution with mean 0 and variance $\sigma^2$, where the variance of the noise depends on $\ell_2$-sensitivity of the projection matrix $P$ and the privacy parameters $\epsilon$ and $\delta$. By choosing $\sigma$ satisfying the condition in Theorem 1, the algorithm guarantees $(\epsilon, \delta)$-differential privacy.

### 3.1.5 Recover distance algorithm

We next describe our algorithm for estimating the squared distance between two users, given their sketches released in a privacy-preserving manner using Algorithm 1. Algorithm 2 (RECOVERDISTANCEPP) computes the squared $\ell_2$ distance between the transformed representations in the $k$ dimensional space, and then discounts for the systemic positive distortion of the distance due to noise addition. The discount $2k\sigma^2$ represents the expected distortion in the squared distance due to Gaussian noise addition.

By repeated application of Algorithm 2, a third party can perform user segmentation and study the characteristics of the segments, as well as perform nearest-neighbor computations.

---
**Algorithm 2** RECOVERDISTANCEPP
---
**Input:** $n \times k$ matrix $Z$ published in a privacy-preserving manner; Noise parameter $\sigma$; Indices $a$, $b$ of the desired users.
**Output:** Estimated squared distance between users $a$ and $b$ in the original space.
1: Let $\hat{x}$ and $\hat{y}$ be the $a$th and $b$th rows in $Z$, respectively.
2: Output $\text{dist}^2_{\text{PP}}(\text{user}_a, \text{user}_b) = \|\hat{x} - \hat{y}\|_2^2 - 2k\sigma^2$.

## 3.2 Formal privacy and utility guarantees

We now prove formal privacy and utility guarantees for the blueprints of Algorithm 1 and Algorithm 2, for the case when noise $\Delta$ is drawn from Normal distribution and the utility goal is to preserve $\ell_2$ distance between users.

### 3.2.1 Privacy guarantees

As Algorithm 2 uses already published data, it is sufficient to provide privacy guarantees for Algorithm 1.

**Theorem 1.** *Let $w_2(P)$ be the $\ell_2$-sensitivity of the projection matrix $P$ (see Definition 2). Assuming $\delta < \frac{1}{2}$, let the entries of the noise matrix be drawn from $N(0, \sigma^2)$ with $\sigma \geq w_2(P)\frac{\sqrt{2(\ln(\frac{1}{2\delta})+\epsilon)}}{\epsilon}$. Then Algorithm 1 satisfies $(\epsilon, \delta)$-differential privacy wrt a change in an individual person's attribute.*

A surprising feature of the algorithm and one that will turn out to be crucial for the utility of the algorithm is that the amount of noise one needs to add in order to satisfy privacy guarantees does not depend on the dimensions of the projection matrix $P$ other than through a (possible) dependence of sensitivity $w_2(P)$ on $P$'s dimension. The work of McSherry and Mironov [24] uses a similar observation relating multi-dimensional Gaussian noise and privacy guarantees without detailing the proof, so we provide the proof for completeness.

We first prove a more general geometric statement, which we will then use to prove the privacy guarantees of our algorithm. The lemma extends the result of Dwork et al. [11] to multiple dimensions.

**Lemma 1.** *Let $Y$ and $Y'$ be points in $\mathbb{R}^l$ s.t. $\|Y - Y'\|_2 \leq w$. Then for any $\hat{D} \subset \mathbb{R}^l$, and any $\Delta$ drawn from $N^l(0, \sigma^2)$, where $\sigma \geq w\frac{\sqrt{2(\ln(\frac{1}{2\delta})+\epsilon)}}{\epsilon}$ and $\delta < \frac{1}{2}$, the following inequality holds: $\Pr[Y' + \Delta \in \hat{D}] \leq e^\epsilon \Pr[Y + \Delta \in \hat{D}] + \delta$.*

*Proof.* The crucial insight is that due to spherical symmetry properties of Gaussian noise, we may choose the basis in such a way that $Y$ and $Y'$ differ in exactly one dimension.

Partition $\hat{D}$ into two sets of points: $\hat{D}_{\text{in}} = \{D \in \hat{D} \colon \langle Y' - Y, D - Y' \rangle \leq wR\}$ and $\hat{D}_{\text{out}} = \{D \in \hat{D} \colon \langle Y' - Y, D - Y' \rangle > wR\}$. The value of $R$ will be determined later. We first prove that

$$\Pr[Y' + \Delta \in \hat{D}_{\text{in}}] \leq e^\epsilon \Pr[Y + \Delta \in \hat{D}_{\text{in}}], \text{ if } R \geq \frac{2\epsilon\sigma^2 - w^2}{2w}, \quad (1)$$

and then prove that

$$\Pr[Y' + \Delta \in \hat{D}_{\text{out}}] \leq \delta, \text{ if } R \geq \sigma\sqrt{2\ln(\frac{1}{2\delta})}. \quad (2)$$

By choosing $\sigma$ so that $R$ satisfies both constraints of (1) and (2), summing the resulting inequalities, and observing that $\Pr[Y + \Delta \in \hat{D}_{\text{in}}] \leq \Pr[Y + \Delta \in \hat{D}]$, we will obtain the desired bound.

**Proof of (1).** By assumption $R \geq \frac{2\epsilon\sigma^2 - w^2}{2w}$. By definition of the Gaussian noise

$$\Pr[Y' + \Delta \in \hat{D}_{\text{in}}] = \frac{1}{(\sqrt{2\pi}\sigma)^k} \int_{\hat{D}_{\text{in}}} \exp\left(-\frac{1}{2\sigma^2}\|Y' - z\|_2^2\right) dz.$$

The density function restricted to $\hat{D}_{\text{in}}$ satisfies:

$$\begin{aligned}
\exp\left(-\frac{1}{2\sigma^2}\|Y'-z\|_2^2\right) &= \exp\left(-\frac{1}{2\sigma^2}\left(\|Y-z\|_2^2 - \|Y-Y'\|_2^2 - 2\langle Y'-Y, z-Y'\rangle\right)\right) \\
&= \exp\left(-\frac{1}{2\sigma^2}\|Y-z\|_2^2\right) \cdot \exp\left(\frac{1}{2\sigma^2}\left(\|Y-Y'\|_2^2 + 2\langle Y'-Y, z-Y'\rangle\right)\right) \\
&\leq \exp\left(-\frac{1}{2\sigma^2}\|Y-z\|_2^2\right) \cdot \exp\left(\frac{w^2 + 2wR}{2\sigma^2}\right) \\
&\leq \exp\left(-\frac{1}{2\sigma^2}\|Y-z\|_2^2\right) \cdot \exp(\epsilon).
\end{aligned}$$

It implies that

$$\begin{aligned}
\Pr[Y' + \Delta \in \hat{D}_{\text{in}}] &= \frac{1}{(\sqrt{2\pi}\sigma)^k} \int_{\hat{D}_{\text{in}}} \exp\left(-\frac{1}{2\sigma^2}\|Y'-z\|_2^2\right) dz \\
&\leq \frac{1}{(\sqrt{2\pi}\sigma)^k} \int_{\hat{D}_{\text{in}}} \exp\left(-\frac{1}{2\sigma^2}\|Y-z\|_2^2\right) \exp(\epsilon) \, dz \leq \exp(\epsilon) \Pr[Y + \Delta \in \hat{D}_{\text{in}}].
\end{aligned}$$

**Proof of (2).** Recall that $R \geq \sigma\sqrt{2\ln(\frac{1}{2\delta})}$. We choose the coordinate system so that $Y = (y_1, \ldots, y_k)$ and $Y' = (y'_1, \ldots, y'_k)$ differ only in the first coordinate and $y'_1 < y_1$. Then

$$\hat{D}_{\text{out}} = \{D \in \hat{D} : \langle Y'-Y, D-Y'\rangle > wR\} \subseteq \{z \in \mathbb{R}^k : (y'_1 - y_1)(z_1 - y'_1) > wR\},$$

which implies the following bound on the probability of $Y' + \Delta$ falling inside $\hat{D}_{\text{out}}$:

$$\begin{aligned}
\Pr[Y' + \Delta \in \hat{D}_{\text{out}}] &= \frac{1}{(\sqrt{2\pi}\sigma)^k} \int_{\hat{D}_{\text{out}}} \exp\left(-\frac{1}{2\sigma^2}\|Y' - z\|_2^2\right) dz \\
&\leq \frac{1}{(\sqrt{2\pi}\sigma)^k} \int_{((y'_1-y_1)(z_1-y'_1)>wR)} \int_{-\infty}^{+\infty} \cdots \int_{-\infty}^{+\infty} \exp\left(-\frac{1}{2\sigma^2}\|Y' - z\|_2^2\right) dz_1 \ldots dz_k \\
&= \frac{1}{(\sqrt{2\pi}\sigma)^k} \int_{(y'_1-y_1)(z_1-y'_1)>wR} \int_{-\infty}^{+\infty} \cdots \int_{-\infty}^{+\infty} \prod_{i=1}^{k} \exp\left(-\frac{1}{2\sigma^2}(y'_i - z_i)^2\right) dz_1 \ldots dz_k \\
&= \frac{1}{\sqrt{2\pi}\sigma} \int_{(y'_1-y_1)(z_1-y'_1)>wR} \exp\left(-\frac{1}{2\sigma^2}(y'_1 - z_1)^2\right) dz_1 \\
&= \frac{1}{\sqrt{2\pi}\sigma} \int_{-\infty}^{\frac{wR}{y'_1-y_1}+y'_1} \exp\left(-\frac{1}{2\sigma^2}(y'_1 - z_1)^2\right) dz_1 \\
&= \frac{1}{2}\left(1 + \operatorname{erf}\left(\frac{\frac{wR}{y'_1-y_1} + y'_1 - y'_1}{\sqrt{2}\sigma}\right)\right) \\
&= \frac{1}{2}\left(1 - \operatorname{erf}\left(\frac{wR}{\sqrt{2}\sigma(y_1-y'_1)}\right)\right) \\
&\leq \frac{1}{2}\exp\left(-\left(\frac{wR}{\sqrt{2}\sigma(y_1-y'_1)}\right)^2\right) \quad (*) \\
&\leq \frac{1}{2}\exp\left(-\frac{w^2R^2}{2\sigma^2 w^2}\right) = \frac{1}{2}\exp\left(-\frac{R^2}{2\sigma^2}\right) \\
&\leq \delta,
\end{aligned}$$

if $R \geq \sigma\sqrt{2\ln(\frac{1}{2\delta})}$. The bound $(*)$ follows from $1 - \operatorname{erf}(x) \leq \exp(-x^2)$ for $x > 0$ [8].

Hence, for (1) and (2) to hold simultaneously, we need

$$\sigma\sqrt{2\ln(\frac{1}{2\delta})} \leq R \leq \frac{2\epsilon\sigma^2 - w^2}{2w} \text{ and } R > 0.$$

By solving the resulting quadratic inequality we conclude that Lemma 1 holds if $\sigma \geq w\frac{\sqrt{\ln(\frac{1}{2\delta})}+\sqrt{\ln(\frac{1}{2\delta})+\epsilon}}{\sqrt{2\epsilon}}$ and $\delta < \frac{1}{2}$. The claim follows by observing that $\sqrt{2(\ln(\frac{1}{2\delta}) + \epsilon)} > \sqrt{\ln(\frac{1}{2\delta})} + \sqrt{\ln(\frac{1}{2\delta}) + \epsilon}$. □

*Proof of Theorem 1.* The intuition behind the proof is to observe that a one-element difference in matrices $X$ and $X'$ will affect only one row of the projection.

To prove that Algorithm 1 satisfies $(\epsilon, \delta)$-differential privacy, we need to prove that for any two input matrices $X$ and $X'$, which differ in one element $x_{aj}$ (corresponding to user $a$ having 1 or 0 value for attribute $j$), and for any $\hat{D}$, where $\hat{D}$ is a set of possible outputs of the algorithm, namely a set of $n \times k$ matrices, the following inequality holds over the random choices of the algorithm:

$$\Pr[X'P + \Delta \in \hat{D}] \leq e^\epsilon \Pr[XP + \Delta \in \hat{D}] + \delta,$$

where $\Delta$ is a $n \times k$ noise matrix, in which each element is drawn independently at random from $N(0, \sigma^2)$.

Fix $X$ and $X'$, and recall the notation of Algorithm 1. Wlog view $Y$ and $Y'$ (in a natural way) as flattened vectors of length $nk$ rather than $n \times k$ matrices. Observe that if $X$ and $X'$ are binary and $\|X' - X\|_2 = 1$, then $\|Y' - Y\|_2 = \|X'P - XP\|_2 = \|(X' - X) \cdot P\|_2 \leq \max_{1 \leq i \leq d} \sqrt{\sum_{j=1}^{k} P_{ij}^2} = w_2(P)$. Applying the result of Lemma 1 to $Y$ and $Y'$, we obtain the desired privacy guarantee. □

We remark that Theorem 1 applies even if the input matrix $X$ consists of values in $[0, 1]$ instead of Boolean values.

In Figure 1 we depict the exact relationship between the privacy parameters $\epsilon$ and $\delta$, and the variance of the noise needed, by plotting three curves of feasible $(\epsilon, \delta)$ pairs for three choices of $\sigma$. The chart can be used either to determine legitimate values of $\epsilon$ and $\delta$ for a fixed $\sigma$, or vice versa. Fixing the value $\sigma$ to 1.0 implies $(\epsilon, \delta)$-privacy for all values of $\epsilon, \delta$ in the middle curve. Alternatively, one can fix the values $(\epsilon, \delta)$ to $(1, 0.1)$ and find a noise level $\sigma \approx 1.0$ that passes through the point.

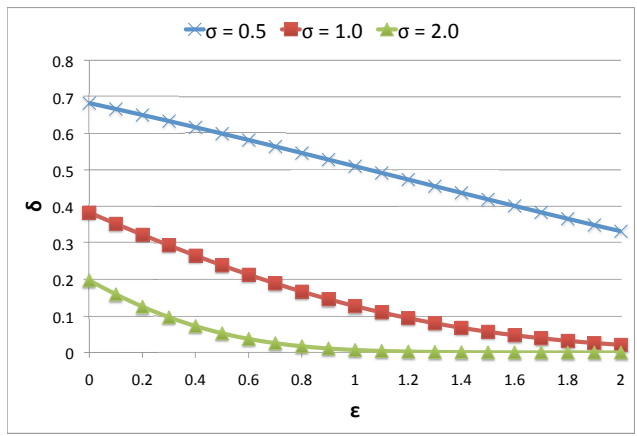

Figure 1: Feasible values of $\epsilon, \delta$ for a given choice of $\sigma$.

### 3.2.2 Utility guarantees for the Gaussian projection

We next discuss the utility guarantees provided by our algorithms. We show that the squared Euclidean distance between two user vectors is preserved in expectation after the privacy transformations performed by our algorithms, and further provide guarantees on how far the distance after transformation can deviate from the original distance. From a third party's perspective, these guarantees imply that (a) the users who are close in the original space are likely to remain close in the transformed space and (b) similarly the users who are far apart are likely to remain so after the transformations.

Concrete utility guarantees depend on the type of the projection matrix $P$. Among the possible choices for projection matrices described in Section 3.1.3, we analyze the guarantees afforded by the use of the Gaussian projection matrix due to Indyk and Motwani [17], proving that the resulting estimate of the squared Euclidean distance is unbiased, computing its variance and giving a tail probability bound.

Although Algorithm 1 is stated as applied to $n$ user vectors simultaneously, we will analyze its utility in preserving squared distances between a particular fixed pair of users. Consider two user vectors $x, y \in \{0, 1\}^d$ which are transformed by Algorithm 1 into $\hat{x} = xP + \Delta_1$ and $\hat{y} = yP + \Delta_2$, where $\Delta_1$ and $\Delta_2$ are independent $k$-dimensional Gaussians $N^k(0, \sigma^2)$.

Recall that according to Theorem 1 in order for Algorithm 1 to satisfy $(\epsilon, \delta)$ differential privacy, $\sigma$ is determined as a function of $\epsilon, \delta$, and $w_2(P)$, which in turn depends on the projection matrix $P$. The following lemma bounds $\sigma$ for a given setting of $\epsilon, \delta$ and $k$.

**Lemma 2.** *Let the projection matrix $P$ be $d \times k$ matrix whose entries are i.i.d. $N(0, 1/k)$ random variables. Algorithm 1 using the noise matrix whose entries are sampled from $N(0, \sigma^2)$ satisfies $(\epsilon, \delta)$-differential privacy if*

$$\sigma \geq \frac{4}{\epsilon}\sqrt{\ln(1/\delta)},$$
$$k > 2(\ln d + \ln(2/\delta)),$$

*and*

$$\epsilon < \ln(1/\delta).$$

*Proof.* According to Theorem 1, $(\epsilon, \delta/2)$-differential privacy is satisfied if

$$\sigma \geq w_2(P)\frac{\sqrt{2(\ln(1/\delta) + \epsilon)}}{\epsilon}, \quad (3)$$

where $w_2(P)$ is $P$'s $\ell_2$-sensitivity. Since the entries of $P$ are distributed as Gaussians, its sensitivity $w_2(P)$ has the following distribution:

$$w_2(P) \sim \sqrt{\max_{1 \leq i \leq d}\left(\sum_{j=1}^{k}|N(0, 1/k)|^2\right)} \sim \sqrt{\max_{1 \leq i \leq d}\frac{1}{k}Z_i},$$

where $Z_i$'s are i.i.d. $\chi_k^2$ variables (i.e., distributed according to the chi-squared distribution with $k$ degrees of freedom). Choosing $x = \ln d + \ln(2/\delta)$ and applying Lemma 4 (see Appendix), we find that

$$\Pr\left[w_2(P) > 1 + \sqrt{\frac{2x}{k}}\right] < \delta/2.$$

Under the assumption that $k > 2(\ln d + \ln(2/\delta))$, the probability that $w_2(P)$ is greater than 2 is less than $\delta/2$. Combining this bound with (3), we find that Algorithm 1 satisfies $(\epsilon, \delta)$-differential privacy for $\epsilon < \ln(1/\delta)$ if

$$\sigma \geq \frac{4}{\epsilon}\sqrt{\ln(1/\delta)} > \frac{2}{\epsilon}\sqrt{2(\ln(1/\delta) + \epsilon)} \text{ and } k > 2(\ln d + \ln(2/\delta)),$$

which completes the proof. □

The proof of Lemma 2 implies that the value of $\sigma$ can be chosen *independent of $P$*. This property is crucial for the following argument, which repeatedly uses independence of the matrix $P$ and the noise $\Delta$ (scaled by $\sigma$).

**Theorem 2.** *Algorithms 1 and 2, where entries of $P$ are sampled from $N(0, 1/k)$, and $\sigma$ is chosen independently of the realization of $P$, satisfy the following utility guarantees:*

1. $\text{dist}^2_{\text{PP}}$ is an unbiased estimator of $\|x-y\|^2_2$:
$$\text{E}[\text{dist}^2_{\text{PP}}(x,y)] = \|x-y\|^2_2.$$

2. Variance of $\text{dist}^2_{\text{PP}}$ is given by the following expression:
$$\text{Var}[\text{dist}^2_{\text{PP}}(x,y)] = 2\|x-y\|^4_2/k + 8\sigma^2\|x-y\|^2_2 + 8\sigma^4 k.$$

3. Deviations are bounded, i.e., with probability $1-(\delta_{\text{JL}} + \delta_{\chi^2} + \delta_N)$, the following holds:
$$\left|\text{dist}^2_{\text{PP}}(x,y) - \|x-y\|^2_2\right| \leq \lambda_{\text{JL}}\|x-y\|^2_2 + 4\sigma^2\sqrt{k}\lambda_{\chi^2} + 4\sigma^2\lambda^2_{\chi^2} + 4\sigma(1+\lambda_{\text{JL}})\lambda_N\|x-y\|_2, \quad (4)$$

when $\lambda_{\text{JL}} < 1/2$, $\delta_{\text{JL}} \geq 2\exp(-k\lambda^2_{\text{JL}}/6)$, $\delta_{\chi^2} \geq 2\exp(-\lambda^2_{\chi^2})$ and $\delta_N \geq \frac{\exp(-\lambda^2_N)}{\lambda_N\sqrt{\pi}}$.

*Proof.* First we note that $\Delta = \Delta_1 - \Delta_2$ is distributed as a $k$-dimensional Gaussian $N^k(0, 2\sigma^2)$. We can express the random variable $\text{dist}^2_{\text{PP}}(x,y)$ as a sum of three random variables $Z_1, Z_2, Z_3$:

$$\text{dist}^2_{\text{PP}}(x,y) = \|\hat{x}-\hat{y}\|^2_2 - 2k\sigma^2 = \|xP + \Delta_1 - yP - \Delta_2\|^2_2 - 2k\sigma^2 =$$
$$= \|(x-y)P + \Delta\|^2_2 - 2k\sigma^2 = \underbrace{\|(x-y)P\|^2_2}_{Z_1} + \underbrace{2\langle(x-y)P, \Delta\rangle}_{Z_2} + \underbrace{\|\Delta\|^2_2 - 2k\sigma^2}_{Z_3}.$$

For the given fixed choice of user vectors $x$ and $y$, let $z = x-y = (z_1, \ldots, z_d)$ and $r = \|x-y\|_2$. Since the entries of $P$ are i.i.d. according to $N(0, 1/k)$, the projection $(x-y)P$ is distributed according to $N^k(0, r^2/k)$. Indeed, the $i$th entry of $(x-y)P$ has the following distribution:

$$\sum_{j=1}^{d} z_j N(0, 1/k) \sim \sum_{j=1}^{d} N(0, z_j^2/k) \sim N(0, \sum_{j=1}^{d} z_j^2/k) \sim N(0, r^2/k).$$

Using the above expression and Lemma 6 we may write the variables $Z_1, Z_2, Z_3$ as follows:

$$Z_1 \sim \left\|N^k(0, \|x-y\|^2_2/k)\right\|^2_2 = r^2 \cdot \chi^2_k/k,$$
$$Z_2 \sim N(0, 8\sigma^2 Z_1),$$
$$Z_3 \sim 2\sigma^2 \chi^2_k - 2k\sigma^2,$$

where $\chi^2_k$ is the chi-squared distribution with $k$ degrees of freedom defined as the distribution of a sum of the squares of $k$ independent $N(0,1)$ random variables.

**Claim 1.** To show that $\text{dist}^2_{\text{PP}}(x,y)$ is an unbiased estimator for $r^2 = \|x-y\|^2_2$ observe that the mean of the chi-squared distribution with $k$ degrees of freedom is $k$. Therefore,

$$\text{E}[Z_1] = \text{E}\left[r^2 \cdot \chi^2_k/k\right] = r^2,$$
$$\text{E}[Z_2] = 0,$$
$$\text{E}[Z_3] = 0,$$

and thus
$$\text{E}[\text{dist}^2_{\text{PP}}(x,y)] = \text{E}[Z_1] + \text{E}[Z_2] + \text{E}[Z_3] = r^2.$$

**Claim 2.** To compute the variance of $\text{dist}^2_{\text{PP}}(x, y)$, express it as

$$\text{Var}\big(\text{dist}^2_{\text{PP}}(x,y)\big) = \text{Var}(Z_1 + Z_2 + Z_3) = \text{E}[(Z_1+Z_2+Z_3)^2] - \big(\text{E}[Z_1+Z_2+Z_3]\big)^2 =$$
$$= \text{E}[Z_1^2 + Z_2^2 + Z_3^2 + 2Z_1Z_2 + 2Z_1Z_3 + 2Z_2Z_3] - \big(\text{E}[Z_1]+\text{E}[Z_2]+\text{E}[Z_3]\big)^2. \quad (5)$$

Recall that by assumption of the theorem, $\sigma$ is chosen independently of $P$, therefore, $(x-y)P$ and $\Delta$ are independent. The expectations of the pairwise products can be evaluated as follows:

$$\text{E}[Z_1 Z_2] = \text{E}\left[\|(x-y)P\|_2^2 \cdot 2\langle(x-y)P, \Delta\rangle\right] = \text{E}\left[2\langle\|(x-y)P\|_2^2 \cdot (x-y)P, \Delta\rangle\right] = 0, \quad \text{(by Lemma 6)}$$
$$\text{E}[Z_1 Z_3] = \text{E}[Z_1]\text{E}[Z_3] = 0, \quad \text{(since } Z_1 \text{ and } Z_3 \text{ are independent)}$$
$$\text{E}[Z_2 Z_3] = \text{E}\left[2\langle(x-y)P, \Delta\rangle \cdot (\|\Delta\|_2^2 - 2k\sigma^2)\right] = \text{E}\left[2\langle(x-y)P, \Delta \cdot (\|\Delta\|_2^2 - 2k\sigma^2)\rangle\right] = 0. \quad \text{(by Lemma 6)}$$

Analyzing the other terms in equation (5), we have

$$\text{E}[Z_1^2] - \text{E}[Z_1]^2 = \text{Var}(Z_1) = \text{Var}(r^2 \cdot \chi_k^2/k) = \frac{r^4}{k^2}\text{Var}(\chi_k^2) = \frac{r^4}{k^2}2k = \frac{2r^4}{k},$$
$$\text{E}[Z_3^2] - \text{E}[Z_3]^2 = \text{Var}(Z_3) = \text{Var}(2\sigma^2\chi_k^2 - 2k\sigma^2) = 4\sigma^4\text{Var}(\chi_k^2) = 8\sigma^4 k,$$

since $\text{Var}(\chi_k^2) = 2k$.

To finish the computation, we need to evaluate $\text{E}[Z_2^2]$. Recall that $Z_2 = 2\langle(x-y)P, \Delta\rangle$, where $(x-y)P \sim N^k(0, r^2/k)$ and $\Delta \sim N^k(0, 2\sigma^2)$. Since $\text{E}[Z_2] = 0$, the second moment of $Z_2$ is $\text{Var}(Z_2)$, which can be computed as follows:

$$\text{Var}(Z_2) = \text{Var}\left(2\sum_{i=1}^k N(0, r^2/k) \cdot N(0, 2\sigma^2)\right) = k\text{Var}\left(2N(0, r^2/k) \cdot N(0, 2\sigma^2)\right) = 8r^2\sigma^2,$$

(the last equation is because the mean of both Gaussians is zero, in which case the variance of the product of two independent variables is the product of their variances).

Putting the above expressions together into equation (5) we obtain

$$\text{Var}(\text{dist}^2_{\text{PP}}(x,y)) = 2r^4/k + 8\sigma^2 r^2 + 8\sigma^4 k,$$

as claimed.

**Claim 3.** Towards proving deviation bounds, observe that

$$|Z_1 - r^2| < \lambda_{\text{JL}} r^2 \text{ with probability at least } 1 - \delta_{\text{JL}} \quad \text{(by [1, Lemma 4.1])}$$
$$|Z_2| \leq 4\sigma\lambda_N\sqrt{Z_1} \text{ with probability at least } 1 - \delta_N \quad \text{(Lemma 5 in Appendix)}$$
$$|Z_3| \leq 4\sigma^2\sqrt{k}\lambda_{\chi^2} + 4\sigma^2\lambda_{\chi^2}^2 \text{ with probability at least } 1 - \delta_{\chi^2}. \quad \text{(by [21, Lemma 1])}$$

Using the union bound and plugging in the bound on $Z_1$ into the second expression, we obtain the desired bound.

By [1, Lemma 4.1] $|Z_1 - r^2| < \lambda_{\text{JL}} r^2$ with probability at least

$$1 - 2\exp(-\frac{k}{2}(\lambda_{\text{JL}}^2/2 - \lambda_{\text{JL}}^2/3)) > 1 - 2\exp(-k\lambda_{\text{JL}}^2/6) \geq 1 - \delta_{\text{JL}},$$

if $\lambda_{\text{JL}} < 1/2$ and $\delta_{\text{JL}} \geq 2\exp(-k\lambda_{\text{JL}}^2/6)$. $\square$

**Optimal projection dimension:** A natural question that our analysis leaves open is how to find an optimum number of dimensions $k$ to which we should project.

- To find the asymptotic of the optimal target dimension $k$ for a fixed setting of the noise $\sigma$ and the failure probability $\delta_{\text{JL}} + \delta_{\chi^2} + \delta_N$, we equate the failure probabilities $\mu = \delta_{\text{JL}} = \delta_{\chi^2} = \delta_N$ for some fixed $\mu < 1/3$. From the conditions on the $\lambda$'s in the statement of Theorem 2 it follows that $\lambda_{\chi^2} = \Theta(\sqrt{\log 1/\mu})$, $\lambda_N = \Theta(\sqrt{\log 1/\mu})$, and $\lambda_{\text{JL}} = \Theta(\sqrt{\frac{\log 1/\mu}{k}})$. Optimizing the upper bound (4) for $k$ we obtain that $k_{\text{OPT}} = \Theta(\sqrt{\log 1/\mu} \cdot \|x - y\|_2^2 / \sigma^2)$.

- Another approach for finding the optimal target dimension $k$ for a fixed setting of the noise $\sigma$ would be to aim to minimize the variance of the squared distance estimate returned by the algorithm, which happens for $k_{\text{OPT}} = \frac{\|x-y\|_2^2}{2\sigma^2}$.

Both of these analytic approaches imply that the optimal value for the target dimension of the privacy-preserving Johnson-Lindenstrauss transform depends on the expected distance between vectors measured using this mechanism, and it scales inversely proportionally to $\sigma^2 = \Theta(\ln(1/\delta)/\epsilon^2)$ (Lemma 2). For this choice of the parameters, the (additive) error in measuring $\|x - y\|_2^2$ is $O(\sigma \sqrt{\log 1/\mu} \cdot \|x - y\|_2)$ and holds with probability $1 - \mu$ assuming that $\log 1/\mu \ll k$. The variance of the estimator when $k = \|x - y\|_2^2 / (2\sigma^2)$ is $16\sigma^2 \cdot \|x - y\|_2^2$. An algorithm designer applying this algorithm in practice could consider using different projection matrices with varying $k$'s each optimized for a particular range of distances, and would need a logarithmic (in terms of possible distances) number of such projections.

## 4 Alternative Approaches

In this section we consider alternative approaches to release of pairwise distances of $n$ vectors in $\mathbb{R}^d$. The first approach is based on output perturbation, where the noise is added directly to the final outcome of the mechanism, i.e., the $n \times n$ matrix of all pairwise distances. We argue that this method is inferior to privacy-preserving projections (previous section) in most settings. The other method is based on input perturbation, where the noise is added to the raw $d$-dimensional vectors. We compare the method to privacy-preserving projections, and discuss their ranges of applicability.

### 4.1 Direct Noise Addition

A classic result in differential privacy [11] shows that any function can be computed with $(\epsilon, \delta)$-differential privacy as long as the Gaussian noise calibrated according to the $\ell_2$-sensitivity of that function is added to the true function value prior to its announcement (Lemma 1). Thus, a natural alternative approach to the Johnson-Lindenstrauss transform-based algorithm that we proposed is an algorithm publishing noisy versions of pairwise distances between points by adding properly calibrated noise to the true distances. We formalize this approach in Algorithms 3 and 4.

**Algorithm 3** NOISEADDITION

**Input:** Boolean $n \times d$ matrix $X$ whose rows correspond to people and columns correspond to attributes learned about the users by first parties; Privacy parameters $\epsilon, \delta$.
**Output:** Privacy-preserving strictly upper triangular $n \times n$ matrix $Z$, whose entries correspond to noisy pairwise distances squared, which can be published.

1: Construct random $n \times n$ strictly upper triangular noise matrix $\Delta$, based on privacy parameters $\epsilon, \delta$.
2: Let $Y$ be a strictly upper triangular $n \times n$ matrix, such that for $1 \leq i < n, i < j \leq n$, $y_{i,j} = \|x_i - x_j\|_2^2$.
3: $Z := Y + \Delta$
4: Publish $Z$.

---

**Algorithm 4** RECOVERDISTANCENA

**Input:** $n \times n$ matrix $Z$ published in a privacy-preserving manner; Noise parameter $\sigma$; Indices $a$, $b$ of the desired users (assume $a < b$ wlog).
**Output:** Estimated squared distance between users $a$ and $b$.

1: Output $\text{dist}^2_{\text{NA}}(\text{user}_a, \text{user}_b) = z_{a,b}$.

---

Similarly to the analysis in Section 3.2.1, Algorithm 3 preserves privacy if $\sigma > \sqrt{n \cdot 2(\ln(\frac{1}{2\delta}) + \epsilon)}/\epsilon$ if $\delta < 1/2$, since a change in a single bit of $X$ causes $n$ changes in the matrix $Y$, each of magnitude one.

Following the analysis of the previous section, consider the variance of the estimator $\text{dist}^2_{\text{NA}}$. Since it is obtained by adding Gaussian noise drawn from $N(0, \sigma^2)$, it is exactly $\sigma^2$:

$$\text{Var}(\text{dist}^2_{\text{NA}}) = \sigma^2 = \Theta(n \ln(1/\delta)/\epsilon^2).$$

Notice that the variance of the estimator is linear in the number of users $n$ (i.e., rows of the matrix $X$).

## 4.2 Comparison between PRIVATEPROJECTION and NOISEADDITION

We use variance of the estimators to compare the accuracy of two methods for release of privacy-preserving pairwise distances. Recall that

$$\text{Var}[\text{dist}^2_{\text{PP}}(x, y)] = 2\|x - y\|_2^4/k + 8\sigma_{\text{PP}}^2\|x - y\|_2^2 + 8\sigma_{\text{PP}}^4 k,$$

where $\sigma_{\text{PP}} = \Theta(\sqrt{\ln(1/\delta)}/\epsilon)$ and $k$ is the target dimension of the projection matrix. The distance $\|x - y\|_2$ for binary vectors $x$ and $y$ is equal to their Hamming distances, and does not exceed the sum of their weights. Let the maximal weight of a user vector be $\mu$. For the target dimension $k = \Theta(\mu)$ (notice that it does not have to be optimal for the the given $\|x - y\|_2^2$):

$$\text{Var}[\text{dist}^2_{\text{PP}}(x, y)] = 2\|x - y\|_2^4/k + 8\sigma_{\text{PP}}^2\|x - y\|_2^2 + 8\sigma_{\text{PP}}^4 k = O(\sigma_{\text{PP}}^4 \cdot \mu).$$

As argued in the previous section,

$$\text{Var}(\text{dist}^2_{\text{NA}}) = \Theta(n \ln(1/\delta)/\epsilon^2),$$

independent of $x$ and $y$. We see that the the variance of the NOISEADDITION method is larger than the variance of PRIVATEPROJECTION if $\mu \sigma_{\text{PP}}^2 = o(n)$. In other words, the PRIVATEPROJECTION method is superior (in terms of the variance of the estimator) if the maximal weight of a user vector is much smaller than the total number of users.

## 4.3 Randomized response

We next describe a technique known as randomized response studied by Warner in the 1960s [32]. Randomized response is a natural, alternate solution to computing privacy-preserving user sketches. We compare this solution to our PRIVATEPROJECTION (PP) approach.

As before, we describe two separate algorithms: one technique for publishing data in a way that preserves privacy and another technique for estimating the squared $\ell_2$ distance between the original vectors given only the perturbed, private vectors. We show that randomized response's strength is in preserving large distances between users, whereas the strength of PP is in preserving small distances. Given the potential applications that we consider, of user segmentation and finding users near a given user, we conclude that PP is a more favorable solution.

### 4.3.1 Privacy guarantees

The algorithm suggested in the randomized response literature for preserving privacy is quite simple: Each bit of a user's vector is flipped with probability $p$ (Algorithm 5). Observe that if $p = \frac{1}{2}$ then the technique achieves perfect privacy, since any vector is equally likely to be published. However, publishing a random vector is worthless. On the other hand, if each bit is flipped with probability slightly less than $\frac{1}{2}$, as in Algorithm 5, then one can show that some privacy is still preserved and yet the perturbed vectors can still be used to estimate the actual distance between vectors.

---
**Algorithm 5** RANDOMIZEDRESPONSE

**Input:** Boolean $n \times d$ matrix $X$ whose rows correspond to people and columns correspond to attributes; Privacy parameter $p < \frac{1}{2}$.
**Output:** Privacy-preserving $n \times d$ matrix $\hat{X}$.

1: $\hat{X}_{ij} := \begin{cases} X_{ij} & \text{with probability } 1-p \\ \overline{X_{ij}} & \text{with probability } p \end{cases}$.
2: Publish $\hat{X}$.

---

We discuss the relationship between the flipping probability $p$ and differential privacy first.

**Lemma 3.** *Algorithm 5 preserves $(\epsilon, 0)$-differential privacy when $\log \frac{1-p}{p} \leq \epsilon$, or equivalently when $p \geq \frac{1}{1+e^\epsilon}$.*

The proof [26, 32] follows by considering two candidate vectors $x$ and $x'$ that differ in only one bit position, and showing that the ratio of the probability that $\hat{x}$ is published given $x$ to the probability that $\hat{x}$ is published given $x'$ is at most $\frac{1-p}{p}$. Setting this value to at most $e^\epsilon$ per the definition of differential privacy yields the lemma.

### 4.3.2 Utility guarantees

We now demonstrate that a third party equipped with the perturbed, private vectors published by Algorithm 5 can still approximate the squared $\ell_2$ distance between pairs of users, via Algorithm 6. The algorithm first computes the squared $\ell_2$ distance between the perturbed representations, and then accounts for the systemic distortion due to perturbation.

**Algorithm 6** RECOVERDISTANCERR

**Input:** $n \times d$ matrix $\hat{X}$ published in a privacy-preserving manner using Algorithm 5; Privacy parameter $p$ used; Indices $a$ and $b$ of the desired users.
**Output:** Estimated squared distance between users $a$ and $b$ before perturbation.
1: Let $x$ and $y$ represent vectors corresponding to users $a$ and $b$ before randomized response perturbation, and let $\hat{x}$ and $\hat{y}$ be the corresponding vectors after perturbation in the published matrix $\hat{X}$.
2: Output $\text{dist}^2_{\text{RR}}(\text{user}_a, \text{user}_b) = \frac{\|\hat{x}-\hat{y}\|_2^2 - 2dp(1-p)}{(1-2p)^2}$

---

**Theorem 3.** *Algorithms 5 and 6 satisfy the following utility guarantees:*
1. $\text{dist}^2_{\text{RR}}$ *is an unbiased estimator of* $\|x-y\|_2^2$:

$$\text{E}[\text{dist}^2_{\text{RR}}(x,y)] = \|x-y\|_2^2.$$

2. *Deviations on squared distances are bounded as follows:*

$$\left|\text{dist}^2_{\text{RR}}(x,y) - \|x-y\|_2^2\right| \leq \frac{\sqrt{d\log(2/\delta_{\text{RR}})}}{\sqrt{2}(1-2p)^2}$$

*with probability at least* $1 - \delta_{RR}$.

*Proof of Theorem 3.* Let $x$ and $y$ be two vectors which after going through the randomized response process yield perturbed vectors $\hat{x}$ and $\hat{y}$. Let $w = \|x-y\|_2^2$. We prove that $\text{dist}^2_{\text{RR}}(y,x)$ is an unbiased estimate for $w$ and is tightly concentrated around $w$.

**Claim 1.** Assume wlog that $x$ and $y$ differ in the first $w$ bits and agree on the remaining $d-w$ bits. In the first $w$ bits, $\text{E}[\|\hat{x}-\hat{y}\|_2^2]$ is the expected number of positions where neither $x$ nor $y$ get flipped or both get flipped. In the remaining $d-w$ positions, $\text{E}[\|\hat{x}-\hat{y}\|_2^2]$ is the expected number of positions where one gets flipped and not the other. Consequently, $\text{E}[\|\hat{x}-\hat{y}\|_2^2] = ((1-p)^2 + p^2)w + 2p(1-p)(d-w) = (1-2p)^2 w + 2p(1-p)d$. Thus

$$\text{E}[\text{dist}_{\text{RR}}(x,y)] = \frac{\text{E}[\|\hat{x}-\hat{y}\|_2^2] - 2dp(1-p)}{(1-2p)^2} = w = \|y-x\|_2^2.$$

**Claim 2.** Observe that for any two bit values, the probability that the distance between them remains unchanged is $q = p^2 + (1-p)^2$, corresponding to both bits either being flipped or both remaining unchanged. Accordingly, the probability that the distance between any two bits changes is $1-q$.

Let $I_i$ denote the indicator random variable corresponding to the distance between $i$th bit of $x$ and $y$ remaining unchanged despite perturbation. Then each $I_i$ can be viewed as an independent Bernoulli trial, with $\Pr[I_i = 1] = q$.

Let $a = \sum_{i=1}^{w} I_i$, and $\overline{b} = \sum_{i=w+1}^{d} \overline{I}_i$. In other words, let $a$ be the number of bit positions among the first $w$ bits in which the distance between bits remains unchanged, i.e., remains 1, and let $\overline{b}$ be the number of bit positions among the remaining $d-w$ bits, where the distance between bits changed (i.e., increases to 1), due to perturbation introduced by Algorithm 5. Then $\|\hat{x}-\hat{y}\|_2^2 = a + \overline{b}$.

By Hoeffding's inequality (applied to $d$ independent random variables, the variance of each of which is bounded):

$$\Pr\left[\|\hat{x}-\hat{y}\|_2^2 - \mathrm{E}[\|\hat{x}-\hat{y}\|_2^2]| \geq \gamma\right] = \Pr\left[|a+\overline{b} - \mathrm{E}[a+\overline{b}]| \geq \gamma\right] \leq 2\exp(-\frac{2\gamma^2}{d}).$$

Therefore,

$$\Pr\left[\left|\mathrm{dist}_{\mathrm{RR}}^2(y,x) - \|y-x\|_2^2\right| \geq \gamma\right] = \Pr\left[\left|\frac{\|\hat{y}-\hat{x}\|_2^2 - 2dp(1-p)}{(1-2p)^2} - \|y-x\|_2^2\right| \geq \gamma\right]$$

$$= \Pr\left[\left|\frac{\|\hat{y}-\hat{x}\|_2^2 + (1-2p)^2 w - \mathrm{E}[\|\hat{y}-\hat{x}\|_2^2]}{(1-2p)^2} - \|y-x\|_2^2\right| \geq \gamma\right]$$

$$= \Pr\left[|\|\hat{y}-\hat{x}\|_2^2 - \mathrm{E}[\|\hat{y}-\hat{x}\|_2^2]| \geq \gamma(1-2p)^2\right]$$

$$\leq 2\exp\left(-\frac{2\gamma^2(1-2p)^4}{d}\right).$$

Plugging in $\gamma = \sqrt{\frac{d\log(\frac{2}{\delta_{RR}})}{2(1-2p)^4}}$, we obtain the desired inequality. $\square$

### 4.3.3 Comparison between PRIVATEPROJECTION and RANDOMIZEDRESPONSE

In Theorems 2 and 3, we showed that both PRIVATEPROJECTION (PP) and RANDOMIZEDRESPONSE (RR) algorithms preserve the expected squared distance between pairs of users, and computed the bounds on how likely it is that the actual values are concentrated around the expectation.

Since both concentration bounds are known to be tight in practice (see Venkatasubramanian and Wang [31] for an empirical study of the Johnson-Lindenstrauss transform), we follow the standard practice of comparing the concentration guarantees to determine which of the two privacy-preserving algorithms would better preserve utility.

Consider the case when $k$ is fixed. When the squared distance is $O(\sqrt{k})$, we show that both algorithms are inaccurate, when the squared distance is between $\sqrt{k}$ and $\sqrt{dk}/\epsilon^2$ our PRIVATEPROJECTION algorithm is more accurate, and when the squared distance is larger than $\sqrt{dk}/\epsilon^2$, RANDOMIZEDRESPONSE is preferable. To see why, consider that it follows from Lemma 3 that for Algorithm 5 to satisfy $(\epsilon, 0)$-differential privacy, the flip probability $p$ has to be such that $p \geq \frac{1}{1+e^\epsilon} \approx 1/2 - \epsilon/4$, which is accurate to within 10% for $\epsilon < 1$. For the purpose of comparison we choose $\sigma = 2\sqrt{\ln(n)}/\epsilon$, resulting in $(\epsilon, 1/n)$-differential privacy according to Theorem 1. This is a conservative setting of the privacy parameters, roughly corresponding to a single violation of the $(\epsilon, 0)$-differential privacy guarantee over $n$ users. Then, equating failure probabilities $\mu = \delta_{\mathrm{RR}} = \delta_{\mathrm{JL}} = \delta_N = \delta_\chi^2$ for some $\mu \ll 1$, we have $\lambda_{\mathrm{JL}} = \Theta(\sqrt{\log(1/\mu)}/\sqrt{k})$, $\lambda_{\chi^2} = \Theta(\sqrt{\log(1/\mu)})$, $\lambda_N = \Theta(\sqrt{\log(1/\mu)})$ for some $k$. Fix two vectors $x, y \in \mathbb{R}^d$ and compare the error of the estimates $\mathrm{dist}_{\mathrm{PP}}^2(x,y)$ and $\mathrm{dist}_{\mathrm{RR}}^2(x,y)$ of the true squared $\ell_2$-distance $\|x-y\|_2^2$. As long as $\sqrt{k} < \|x-y\|_2^2 < \sqrt{dk}/\epsilon^2$, which controls the first term of the bound (4), and $k(\ln n)^2 \ll d$, which bounds the second term, the estimate $\mathrm{dist}_{\mathrm{PP}}^2(x,y)$ is closer to the true distance, and hence PRIVATEPROJECTION outperforms RANDOMIZEDRESPONSE. The exact constants separating these regions depend on the privacy parameters and failure probabilities $\delta$'s.

Note however that the target dimension $k$ is not fixed, but rather is selected by the curator. $k$ can be selected with the goal of finding the sweet spot between preserving privacy and the utility of a given algorithm.

Alternatively, several sketches with different values of $k$ can be released so as to preserve distances at multiple scales, each consuming its share of the privacy budget.

## 5 Discussion

In this section, we describe how user sketches released in a privacy-preserving way can be used by third parties, and conclude by discussing the limitations of sketches and our privacy guarantees.

### 5.1 Applications

We begin by re-iterating exactly what can be safely published:
1. The $d$ attribute meanings in the original vector space, assuming the meanings themselves are not sensitive, or the ones that are published via a method similar to the one in [20] or [16].
2. The Johnson-Lindenstrauss projection matrix $P$.
3. For each user $x$, their userID, together with their perturbed sketch $xP + \Delta$.

There are several actions that the third party can perform with this published information, depending on what kind of additional information the third party possesses about the users and the goal the third party is trying to achieve.

**Segmentation:** User sketches can be segmented via some clustering algorithm and then information known to the third party about some members of the cluster can be generalized to the rest of the cluster. There is convincing evidence that segmentation of users into clusters is effective in some contexts [4,25,27,33].

**Nearest Neighbors:** Another application of perturbed sketches is finding nearest neighbors. For example, finding users most similar to an already known one can be useful in the context of online dating, and product and movie recommendations.

### 5.2 Limitations

Although our algorithm offers a method for privacy-preserving sharing of user data with third parties in a way that enables user-user distance computations, there are other tasks for which the user data shared using our method would not be useful to third parties. We also discuss the limitations of the privacy protections we provide.

#### 5.2.1 Utility Limitations

An important limitation of our work from the utility perspective is that the dimensions of the user sketches are impossible to interpret. As a consequence, the only way for a third party to select users satisfying a particular attribute is to project the vector corresponding to this attribute in the higher-dimensional space to the lower-dimensional space, and then perform the distance computation between user sketches and the obtained lower-dimensional attribute vector. However, as explained in §4.3.3, this computation would fall into the range of squared distance values for which both PRIVATEPROJECTION and RANDOMIZEDRESPONSE perform poorly.

Furthermore, the proposed computation of user sketches weighs all attributes equally, which may not be desirable for third parties who want to prioritize similarity between users in some of the attributes over others. Computing multiple projections, each based on a different subset or weighing of the attributes would require use of additional privacy budget for each projection, as well as necessitate precluding the possibility of collusion among third parties.

Another limitation, which is a challenge for much of the privacy literature, is that our sketches provide a static snapshot of user data, and would require additional privacy budget in order to update them as the user information changes. The work of [13] offers directions for possibly overcoming this challenge.

### 5.2.2 Privacy Limitations

While our work takes an important step forward, privacy is more complex than ensuring that a third party cannot infer a particular attribute of a user. For example, if many of the attributes are correlated or representative of a higher-level user feature, then our techniques do not prevent a third party from inferring that. In other words, our guarantees apply to a constant number of attributes, but not to a persistent trend that exists in the data. Depending on the context, it may be more powerful to first categorize the attributes into a coarser granularity prior to producing perturbed sketches.

Finally, as we explained in the Introduction, the goal of our work is to enable third parties to perform distance computations and clustering on users. Clearly, our work is not relevant for settings where such computations and privacy are fundamentally at odds, i.e., scenarios where the underlying data is so sensitive that even the ability to identify that two users are similar constitutes a privacy violation.

## 6 Related Work

Liu et al. [22] introduce and motivate the problem of releasing data to third parties with a goal that the original sensitive information cannot be inferred while preserving analytic properties of the data, such as inner product and Euclidean distance computations. Their approach is based on random projection to a lower-dimensional subspace using a projection matrix drawn from a distribution unknown to the adversary. The key distinction from our work is that they do not utilize an operational definition for what it means to protect the privacy of the data, and therefore, as they point out, there are scenarios in which an adversary can find approximations to original data (e.g., if the data is restricted to Boolean domain or adversary possesses certain background knowledge). Follow-up works [14, 30] propose concrete attacks and demonstrate the vulnerabilities of the approach. Our use of differential privacy and addition of properly calibrated random noise after the projection enables us to provide a rigorous privacy guarantee, as well as gain insight into the change in utility depending on projected dimension used. Mukherjee et al. [28] propose enabling distance-based mining algorithms over private data using Fourier-related transforms, but their approach has the same drawbacks as Liu et al. [22].

We discussed randomized response in §4.3 and compared its performance with our method in §4.3.3. In terms of privacy, randomized response offers a slightly better privacy guarantee. However, from a utility perspective, randomized response does not preserve "small" distances as effectively as the present work. Concretely, for users that are less than $\sqrt{dk}$ distance apart, our method provides stronger guarantee than randomized response. Since third parties are likely interested in preserving small distances, we believe that our approach is more suitable for typical data mining applications.

From a differential privacy perspective, alternative solutions could be used to attack our problem. For example, Blum et al. [5] give a method of running $k$-Means on a private data set maintained by a trusted administrator. Their goal is to produce $k$ cluster centers that are not too far from the $k$ cluster centers that $k$-Means would produce if the algorithm had access to the private data. Our goal differs in that we seek to publish data (or enable its utilization) along with the userIDs and enable identification of users who belong to the same cluster. Also, our noisy sketches can also be used for other distance-based computations such as nearest neighbors. Finally, another possible direction is not to publish any data and only allow black-box queries to the first-party data provider, where the answers to such queries are perturbed [12, 23]. This approach may place considerable burden on the first-party data provider, and consumes privacy budget with each query posed.

## 7 Conclusion and Future Work

We proposed a viable solution to the challenge of publishing user data for enabling computation of distance between users, without revealing the values of user data attributes. The key insight behind our technique is that by projecting users to a lower-dimensional space, we can limit the amount of noise we add to each user's data, while also reaping the benefit of preserving distances. We also compared our proposed solution to other candidate solutions, such as directly adding noise to the pairwise distances or adding noise to each attribute of a user, and showed that our method is preferable for potential applications such as user segmentation and nearest neighbor search.

There is ample opportunity to improve upon our results as the problem of privacy-preserving data sharing with third parties is naturally more complex than sharing in a way that enables distance computations. For example, third parties would benefit from data that enables computation of other data-mining primitives and from ability to operate on dynamically changing data [13].

## A  Appendix

**Lemma 4.** *Let $X_1, \ldots, X_n$ are i.i.d. variables drawn from $\chi_k^2$ (chi-squared distribution with k degrees of freedom). For any $x > 0$*

$$\Pr[\sqrt{\max_i X_i} > \sqrt{k} + \sqrt{2x}] < \exp(\ln n - x).$$

*Proof.* We use the bound due to Laurent and Massart [21, Lemma 1] on the tail probability of the chi-squared distribution:

$$\Pr[X \geq k + 2\sqrt{kx} + 2x] \leq \exp(-x),$$

where $X \sim \chi_k^2$. We establish the claim by taking the union bound over $n$ independent variables and observing that

$$\sqrt{k + 2\sqrt{kx} + 2x} < \sqrt{(\sqrt{k} + \sqrt{2x})^2} = \sqrt{k} + \sqrt{2x}.$$

for $x > 0$. □

**Lemma 5.** *Suppose $X$ is drawn from $N(0, \sigma^2)$. Then*

$$\Pr[|X| < x] \geq 1 - \exp\left(-\frac{x^2}{2\sigma^2}\right),$$

$$\Pr[|X| < x] \geq 1 - \frac{2\sigma}{x\sqrt{2\pi}} \exp\left(-\frac{x^2}{2\sigma^2}\right).$$

*The second bound is stronger than the first when $x \geq 0.8\sigma$.*

*Proof.* Expressing the tail probabilities in terms of the CDF of the standard Normal distribution denoted as $\Phi$ we have (the first bound):

$$\Pr[X < -x] = \Phi(-x/\sigma),$$
$$\Pr[X > -x] = 1 - \Phi(x/\sigma),$$

and thus

$$\Pr[|X| < x] = 1 - (\Phi(-x/\sigma) + 1 - \Phi(x/\sigma)) = \Phi(x/\sigma) - \Phi(-x/\sigma)$$
$$= \frac{1}{2}\big(1 + \text{erf}(\frac{x}{\sigma\sqrt{2}}) - 1 - \text{erf}(\frac{-x}{\sigma\sqrt{2}})\big) = \text{erf}(\frac{x}{\sigma\sqrt{2}}).$$

Alternatively (the second bound)

$$\Pr[|X| < x] = 1 - 2\Pr[X > x] \geq 1 - \frac{2\sigma}{x\sqrt{2\pi}} \exp\left(-\frac{x^2}{2\sigma^2}\right).$$

The second bound is be stronger than the first as long as

$$2\sigma \frac{1}{x\sqrt{2\pi}} \leq 1,$$

which holds when $x \geq 0.8\sigma$. □

**Lemma 6.** *Let $X$ be an arbitrary distribution over $\mathbb{R}^k$ and $Y \sim N^k(0, \sigma^2)$, independent of $X$. Then*

$$\langle X, Y \rangle \sim N(0, \|X\|_2^2 \sigma^2).$$

*In particular,*

$$\text{E}[\langle X, Y \rangle] = 0.$$

*Proof.* Let $X = (X_1, \ldots, X_k)$ and $Y = (Y_1, \ldots, Y_k)$. Then

$$\langle X, Y \rangle \sim \sum_{i=1}^k X_i Y_i \sim \sum_{i=1}^k X_i N(0, \sigma^2) \sim \sum_{i=1}^k N(0, X_i^2 \sigma^2) \sim N(0, \sigma^2 \sum_{i=1}^k X_i^2) \sim N(0, \sigma^2 \|X\|_2^2),$$

by scaling and additive properties of the Gaussian distribution. □